# CASCADE MIXING AND THE CP-VIOLATING ANGLE BETA[*]

Boris Kayser

National Science Foundation
4201 Wilson Boulevard
Arlington, VA 22230 USA

**Abstract**

In the decay chain $B_d \to \psi + K \to \psi + (\pi \ell \nu)$, neutral K mixing follows on the heels of neutral B mixing. This "cascade mixing" leads to an interference which probes $\cos 2\beta$, where $\beta$ is one of the three CP-violating phase angles which characterize CP violation in the Standard Model. Widely-discussed future B-system experiments will determine trigonometric functions of these three phase angles, leaving the underlying angles themselves discretely ambiguous. A determination of $\cos 2\beta$ through cascade mixing would eliminate all the discrete ambiguities entirely.

---

Suppose a neutral B, which is a coherent superposition of mass eigenstates, decays to a final state containing a neutral K, which is also such a superposition. Then we can have K mixing following on the heels of B mixing. We refer to this as "cascade mixing".[1] As we shall see, cascade mixing could help us test the Standard Model of CP violation by providing important information which cannot be obtained by the more commonly discussed future B- or K-system CP experiments.

Let us call the mass eigenstates of the $B_d$-$\overline{B}_d$ system $B_{Heavy}$ ($B_H$) and $B_{Light}$ ($B_L$). These states have complex masses

$$\lambda_{Heavy(Light)} = M_B \; (\pm) \; \frac{\Delta M_B}{2} - i\frac{\Gamma_B}{2} , \qquad (1)$$

where $M_B$ is their average mass, $\Delta M_B$ is their mass difference and is defined to be positive, and $\Gamma_B$ is their common width. Similarly, the mass eigenstates of the $K^0$-$\overline{K^0}$ system, $K_{Short}$ ($K_S$) and $K_{Long}$ ($K_L$), have complex masses

$$\mu_{S(L)} = m_K \; (\mp) \; \frac{\Delta m_K}{2} - i\frac{\gamma_{S(L)}}{2} , \qquad (2)$$

where $m_K$ is the average of the $K_S$ and $K_L$ masses, $\Delta m_K \equiv m(K_L) - m(K_S)$ is their positive mass difference, and $\gamma_{S,L}$ are the $K_{S,L}$ widths.

Under special circumstances, the composition of a neutral meson in terms of its mass-eigenstate components can be tuned. For example, consider the kaons produced by a "regenerator". Incident on the regenerator is a pure $K_L$ beam. With some amplitude r, the regenerator introduces into the beam a $K_S$ component. That is, when a kaon emerges from the regenerator, it is in the state

$$|K_r\rangle \propto |K_L\rangle + r|K_S\rangle . \qquad (3)$$

By changing the characteristics of the regenerator, one can change r and thereby change the $K_S$, $K_L$ composition of $K_r$.

Suppose, now, that we produce neutral kaons via the decay

$$B_d \xrightarrow[\text{After } \tau_B]{} \psi + K . \qquad (4)$$

In this decay, the parent is produced as a pure $B_d$ at time $\tau_B \equiv 0$ in its rest frame, and then decays into $\psi + K$ after a proper time $\tau_B$. It is straightforward to show that the K created by this decay is born in the state $|K_{From\ B}\rangle$ given by

$$|K_{\text{From B}}\rangle \propto \left( e^{-i\frac{\Delta M_B}{2}\tau_B} + i\tan\beta\, e^{i\frac{\Delta M_B}{2}\tau_B} \right)|K_S\rangle$$
$$+ \left( e^{i\frac{\Delta M_B}{2}\tau_B} + i\tan\beta\, e^{-i\frac{\Delta M_B}{2}\tau_B} \right)|K_L\rangle \quad . \tag{5}$$

Here,

$$\beta \equiv \arg\left[ -\frac{V_{cd}V_{cb}^*}{V_{td}V_{tb}^*} \right] \quad , \tag{6}$$

where V is the Cabibbo-Kobayashi-Maskawa (CKM) quark mixing matrix. Elements of V occur in the diagrams for $B_d$-$\overline{B}_d$ mixing, $K^0$-$\overline{K}^0$ mixing, and B→ψK decay. As a result, the relative phase β of these elements, which if nonvanishing is CP-violating, occurs in the expression (5) for $|K_{\text{From B}}\rangle$.

From Eq. (5), we see that by selecting events with a particular $\tau_B$, we can tune the $K_S$, $K_L$ composition of $K_{\text{From B}}$, just as we can tune that of the $K_r$ from a regenerator. Both the relative phase and magnitude of the $K_S$ and $K_L$ components of $K_{\text{From B}}$ vary with $\tau_B$. In particular, from the squares of the coefficients of $K_S$ and $K_L$ in Eq. (5), we find that

$$\frac{\text{Prob }(K_S)}{\text{Prob }(K_L)} = \frac{1-\sin 2\beta \sin(\Delta M_B \tau_B)}{1+\sin 2\beta \sin(\Delta M_B \tau_B)} \quad . \tag{7}$$

Here, Prob ($K_S$) is the probability of finding a $K_S$ in $K_{\text{From B}}$, and similarly for Prob ($K_L$).

Existing data allow sin 2β to lie anywhere in the range 0.38 to 0.94.[2] From Eq. (7), we see that, so long as sin 2β is not very close to unity, Prob ($K_S$) and Prob ($K_L$) are of the same order of magnitude.

As time passes after the decay $B_d \xrightarrow[\text{After }\tau_B]{} \psi + K$ produces a neutral kaon, the $K_S$, $K_L$ composition of this kaon changes as a result of K-$\overline{K}$ mixing, and we no longer have the composition given by Eq. (5). When the kaon itself decays, its $K_S$ and $K_L$ pieces contribute coherently. The interference between their contributions can reflect both the B mixing which took place before the B decayed, and the K mixing which took place after it decayed. The B mixing plays a role because it determines the state $|K_{\text{From B}}\rangle$ in which the kaon is born, and the subsequent K mixing plays one because it determines how the $K_S$, $K_L$ composition of the kaon changes after it is born.

If the $K_S$-$K_L$ interference when the kaon decays to some final state f is to be significant, the $K_S$ and $K_L$ contributions to the decay must be comparable. Since, as we have seen, the $K_S$ and $K_L$ components of the kaon wave function are of the same order of magnitude, this means that the decay amplitudes $A(K_S \to f)$ and $A(K_L \to f)$ must be comparable. In the table below, we show the ratio $|A(K_L \to f) / A(K_S \to f)|$ for the common final states.[3)] In the table, $\ell$ is an e or μ.

| Final State f | $\left|\dfrac{A(K_L \to f)}{A(K_S \to f)}\right|$ |
|---|---|
| ππ | 0.002 |
| ππγ | 0.007 |
| π$\ell$ν | 1.00 |
| $\pi^+\pi^-\pi^0$ | ~24 |

From this table, we see that it is in the semileptonic decays $K \to \pi\ell\nu$ that $K_S$-$K_L$ interference will be appreciable.

Let us consider, then, the decay chain

$$B_d \xrightarrow{\text{After } \tau_B} \psi + K \xrightarrow{\text{After } \tau_K} \pi^-\ell^+\nu \quad (8)$$

Here, a $B_d$ decays at proper time $\tau_B$ after its birth, as measured in its rest frame, and its daughter K decays at proper time $\tau_K$ after its own birth, as measured in its own rest frame. In the Standard Model (SM), the final state $\pi^-\ell^+\nu$ can come only from the $K^0$, and not the $\overline{K^0}$, component of the K. (There is, of course, also the alternative final state $\pi^+\ell^-\overline{\nu}$, which can come only from the $\overline{K^0}$ component. For the moment, we consider only the final state $\pi^-\ell^+\nu$.) The amplitude for the decay chain (8) is given by[4)]

$$\text{Amp}\left[B_d \xrightarrow{\text{After } \tau_B} \psi + K \xrightarrow{\text{After } \tau_K} \psi + \left(\pi^-\ell^+\nu\right)\right] =$$

$$= \left\{ \sum_{\substack{M=Heavy,\\Light\\N=S,L}} A(B_d \text{ is } B_M) e^{-i\lambda_M \tau_B} A(B_M \to \psi K_N) e^{-i\mu_N \tau_K} A(K_N \text{ is } K^0) \right\} \times \quad (9)$$

$$\times A\left(K^0 \to \pi^-\ell^+\nu\right)$$

Here, $A(B_d \text{ is } B_M)$ is the amplitude for the initial $B_d$ to be, in particular, the B mass eigenstate $B_M$, $\exp(-i\lambda_M \tau_B)$ is the amplitude for this $B_M$ to propagate for a proper time $\tau_B$, $A(B_M \to \psi K_N)$ is the amplitude for the $B_M$ to decay to a $\psi$ plus the K mass eigenstate $K_N$, $\exp(-i\mu_N \tau_K)$ is the amplitude for this K mass eigenstate to propagate for a proper time $\tau_K$, $A(K_N \text{ is } K^0)$ is the amplitude for the $K_N$ to be, in particular, a $K^0$ (the only component of a K that can decay to $\pi^-\ell^+\nu$), and $A(K^0 \to \pi^-\ell^+\nu)$ is the amplitude for a $K^0$ to decay to $\pi^-\ell^+\nu$. That the amplitude for a B of complex mass $\lambda_M$ to propagate for a proper time $\tau_B$ is $\exp(-i\lambda_M \tau_B)$[4] follows trivially from Schrödinger's equation applied in the rest frame of the B. The amplitude for a K to propagate follows similarly.

An amplitude similar to that of Eq. (9) describes the decay chain where the parent B is born a $\overline{B_d}$, and/or the K decay channel is $\pi^+\ell^-\overline{\nu}$.

It is interesting to compare the amplitude (9) for $B_d \to \psi + K \to \psi + (\pi\ell\nu)$ with the corresponding amplitude for the more familiar process $B_d \to \psi + K_S$. [At the practical level, by "$K_S$" one means here a neutral kaon which decays to $\pi\pi$ (which $K_L$ does only rarely) within roughly one $K_S$ lifetime.] We have

$$\text{Amp}\left[ B_d \xrightarrow[\text{After } \tau_B]{} \psi + K_S \right] = \\ = \sum_{\substack{M=Heavy,\\Light}} A(B_d \text{ is } B_M) e^{-i\lambda_M \tau_B} A(B_M \to \psi K_S) \quad . \quad (10)$$

Here, only one K mass eigenstate, and no nontrivial K propagation or K mixing, is involved. While this amplitude sums only over B mass eigenstates, the amplitude (9) for $B_d \to \psi + K \to \psi + (\pi\ell\nu)$ sums over both B and K mass eigenstates, reflecting the nontrivial role of both B and K mixing in the process it describes.

Inserting the SM values for $A(B_d \text{ is } B_M)$ and $A(B_M \to \psi K_S)$ in Eq. (10), and squaring this amplitude and its analogue for a parent B born as a $\overline{B_d}$, we find the famous result[5]

$$\Gamma\left[ \overset{(-)}{B_d} \xrightarrow[\text{After } \tau_B]{} \psi + K_S \right] \propto e^{-\Gamma_B \tau_B} \left[ 1_{(+)}^{(-)} \sin 2\beta \sin(\Delta M_B \tau_B) \right] \quad . \quad (11)$$

Since $\psi K_S$ is a CP eigenstate, these two decay rates are the rates for CP-mirror-image processes. Thus, the difference between them violates CP invariance. As we see, this difference measures $\sin 2\beta$. Moreover, it does so *cleanly*. That is, beyond their dependence on the CP-violating angle $\beta$, which one would like to determine to test the SM of CP violation, the decay rates (11) depend only on parameters which are already known, $\Gamma_B$ and $\Delta M_B$. There is no dependence on unknown or uncertain quantities.

We now insert in Eq. (9) the SM values for A($B_d$ is $B_M$), A($B_M \to \psi K_N$), and A($K_N$ is $K^0$), neglecting the small violation of CP in the K system. Then, squaring the amplitude of Eq. (9), and its counterparts for a parent B which is born as a $\overline{B}_d$ and/or a daughter K which decays into $\pi^+\ell^-\overline{\nu}$, we find that

$$\Gamma\left[\overset{(-)}{B_d} \xrightarrow[\text{After } \tau_B]{} \psi + K \xrightarrow[\text{After } \tau_K]{} \psi + \left(\pi^{\mp}\ell^{\pm}\nu\right)\right] \propto$$

$$\propto e^{-\Gamma_B \tau_B}\left\{ e^{-\gamma_S \tau_K}\left[1_{(\mp)}^{-}\sin 2\beta \sin(\Delta M_B \tau_B)\right]\right.$$

$$+ e^{-\gamma_L \tau_K}\left[1_{(-)}^{\pm}\sin 2\beta \sin(\Delta M_B \tau_B)\right]$$

$$\overset{\pm}{(\mp)} 2 e^{-\frac{1}{2}(\gamma_S+\gamma_L)\tau_K}\left[\cos(\Delta M_B \tau_B)\cos(\Delta m_K \tau_K)\right.$$

$$\left.\left.+ \cos 2\beta \sin(\Delta M_B \tau_B)\sin(\Delta m_K \tau_K)\right]\right\}$$
(12)

Here, at the head of the third line on the right-hand side, the pair of signs which is not (is) in parentheses corresponds to an initial $B_d$ ($\overline{B}_d$). The upper sign of each of these two pairs is for $K \to \pi^-\ell^+\nu$, and the lower one for $K \to \pi^+\ell^-\overline{\nu}$.

From the way in which the various parts of the decay rates (12) depend on $\gamma_S$ or $\gamma_L$, it is obvious that the first line in the expression for these rates is the contribution of the $K_S$ component of the daughter kaon, the second line is the contribution of the $K_L$ component, and the remaining lines are the $K_S$-$K_L$ interference term.

While the rates (12) for $\overset{(-)}{B_d} \to \psi + K \to \psi + (\pi\ell\nu)$ are not as simple as the rates for $\overset{(-)}{B_d} \to \psi + K_S$, they are every bit as clean. That is, they depend only on the angle $\beta$ one would like to determine, and on the already-known parameters $\Gamma_B$, $\gamma_S$, $\gamma_L$, $\Delta M_B$, and $\Delta m_K$.

To test the SM of CP violation, one would like to determine from B decays the CP-violating angles $\alpha$, $\beta$, and $\gamma$, where $\beta$ is defined by Eq. (6),

$$\alpha \equiv \arg\left[-\frac{V_{td}V_{tb}^*}{V_{ud}V_{ub}^*}\right], \quad (13)$$

and

$$\gamma \equiv \arg\left[-\frac{V_{ud}V_{ub}^*}{V_{cd}V_{cb}^*}\right]. \quad (14)$$

These three angles are the interior angles of what is commonly referred to as the CKM unitarity trangle.[6] To carry out the test of the SM, one would like to determine $\alpha$, $\beta$, and $\gamma$ themselves, and not just trigonometric functions of them such as $\sin 2\beta$, which leave the underlying angles discretely ambiguous. Unfortunately, the rates for $\overset{(-)}{B_d} \to \psi + K_S$, the B decays where CP violation will probably be first sought, measure only $\sin 2\beta$ (see Eq. (11)). In contrast, the rates for $\overset{(-)}{B_d} \to \psi + K \to \psi + (\pi\ell\nu)$ are sensitive to both $\sin 2\beta$ and $\cos 2\beta$ (see Eq. (12)). Thus, once $\sin 2\beta$ (hence $|\cos 2\beta|$) has been determined from $\overset{(-)}{B_d} \to \psi + K_S$, the rates for $\overset{(-)}{B_d} \to \psi + K \to \psi + (\pi\ell\nu)$ could be used to determine the sign of $\cos 2\beta$. This additional information could do much more than partially eliminate the discrete ambiguity in $\beta$. Indeed, if, as hoped, $\sin 2\beta$ will have been found from $\overset{(-)}{B_d} \to \psi + K_S$, $\sin 2\alpha$ from $\overset{(-)}{B_d} \to \pi\pi$, and $\cos 2\gamma$ from $B^\pm \to DK^\pm$, a determination of Sign($\cos 2\beta$) would then suffice to eliminate all discrete ambiguities from $\alpha$, $\beta$, and $\gamma$.[7]

To extract Sign($\cos 2\beta$) from the decay rates for $\overset{(-)}{B_d} \to \psi + K \to \psi + (\pi\ell\nu)$, one would measure them as functions of $\tau_B$ and $\tau_K$, and then compare them with Eqs. (12), taking the known values of $\Gamma_B$, $\gamma_S$, $\gamma_L$, $\Delta M_B$, $\Delta m_K$, and $\sin 2\beta$ as inputs. The ease with which the extraction could be performed would depend, of course, on the event rate. Now, the popular decays $\overset{(-)}{B_d} \to \psi + K_S$ will be detected principally through the $\pi^+\pi^-$ decay mode of $K_S$. Thus, we normalize the decay rate (12) relative to (11) by taking into account the branching ratios for kaon decay to $\pi^\mp\ell^\pm\nu$ and to $\pi^+\pi^-$. Then, integrating over $\tau_K$, we find that if, for example, $\cos 2\beta = \sin 2\beta$, then the $\cos 2\beta$ term in $\Gamma[B_d \to \psi + K \to \psi + (\pi^-\ell^+\nu)]$ contributes an event rate 1/600 that contributed by the $\sin 2\beta$ term in $\Gamma[B_d \to \psi + K_S \to \psi + (\pi^+\pi^-)]$. Hopefully, such an event rate makes the study of the sign of the $\cos 2\beta$ term in $\overset{(-)}{B_d} \to \psi + K \to \psi + (\pi\ell\nu)$ feasible at hadron facilities, although it may not be feasible at $e^+e^-$ B factories. It should be noted that the total number of decays of the type $\overset{(-)}{B_d} \to \psi + K \to \psi + (\pi\ell\nu)$ will actually be comparable to the number of the type $\overset{(-)}{B_d} \to \psi + K_S \to \psi + (\pi^+\pi^-)$. However, while the latter decays will be concentrated at values of $\tau_K \lesssim 1/\gamma_S$, the former decays will be spread out over the much larger region $\tau_K \lesssim 1/\gamma_L$. The $K_S$-$K_L$ interference term in $\overset{(-)}{B_d} \to \psi + K \to \psi + (\pi\ell\nu)$, where $\cos 2\beta$ appears, is only significant when $\tau_K \lesssim 2/\gamma_S$, so that the $K_S$ component of the kaon has not already decayed away. Thus, only the small fraction of all $\overset{(-)}{B_d} \to \psi + K \to \psi + (\pi\ell\nu)$ decays which have $\tau_K \lesssim 2/\gamma_S$ are usable. This is how the event rate from the $\cos 2\beta$ term in $\Gamma[\overset{(-)}{B_d} \to \psi + K \to \psi + (\pi\ell\nu)]$ comes to be much smaller than that from the $\sin 2\beta$ term in $\Gamma[\overset{(-)}{B_d} \to \psi + K_S \to \psi + (\pi^+\pi^-)]$.

To determine the decay times $\tau_B$ and $\tau_K$ in an event of the type $\overset{(-)}{B_d} \to \psi + K \to \psi + (\pi\ell\nu)$, one would measure the B and K pathlengths and energies. The presence of an undetectable neutrino in the final state to which the K decays does not make it impossible to determine the K energy. Indeed, despite the neutrino, one would have a four-constraint fit to the kinematics of the entire decay chain.[8]

Why does $B_d \to \psi + K_S$ probe only $\sin 2\beta$, while $B_d \to \psi + K \to \psi + (\pi \ell \nu)$ probes both $\sin 2\beta$ and $\cos 2\beta$? To answer this question, let us consider Fig. 1. As shown there, a particle born as a pure $B_d$ or a pure $\overline{B}_d$ has a $B_H$ and a $B_L$ component. Either of these mass eigenstate components can decay to either $\psi + K_L$ or $\psi + K_S$. Subsequently, either the $K_L$ or $K_S$ can decay to $\pi \ell \nu$. Thus, as illustrated in Fig. 1, there are four paths through which the parent $\overset{(-)}{B_d}$ can produce the final state $\psi + (\pi \ell \nu)$. Now, in the limit where CP is conserved and $\beta$ vanishes, the intermediate states in Fig. 1, $B_H$, $B_L$, $\psi K_L$, and $\psi K_S$, are all CP eigenstates. In particular, $CP(B_H) = CP(\psi K_S) = -$, while $CP(B_L) = CP(\psi K_L) = +$.[9] Thus, the decays $B_H \to \psi K_L$ and $B_L \to \psi K_S$, represented by dashed lines in Fig. 1, connect states which in the CP-conserving limit are of opposite CP parity. Consequently, the amplitudes for these decays must vanish as CP violation (hence $\beta$) goes to zero, and one finds by explicit calculation that they are proportional to $\sin \beta$. In contrast, the decays $B_H \to \psi K_S$ and $B_L \to \psi K_L$, represented by solid lines in Fig. 1, connect states which in the CP-conserving limit are of the same CP parity. Thus, the amplitudes for these decays are expected to survive as CP violation goes to zero, and one finds that they are proportional to $\cos \beta$. Now, from Fig. 1, we see that the decays $\overset{(-)}{B_d} \to \psi + K_S$ involve only two paths, one through $B_H \to \psi K_S$, and one through $B_L \to \psi K_S$. It is the interference between the amplitudes for these two paths that leads to CP violation in $\overset{(-)}{B_d} \to \psi + K_S$. Since $A(B_H \to \psi K_S) \propto \cos \beta$, while $A(B_L \to \psi K_S) \propto \sin \beta$, this interference is proportional to $\cos \beta \sin \beta$, or $\sin 2\beta$. This is why $\overset{(-)}{B_d} \to \psi + K_S$ probes only $\sin 2\beta$. In contrast, in the decay chain $\overset{(-)}{B_d} \to \psi + K \to \psi + (\pi \ell \nu)$, there are the four paths shown in Fig. 1, and all of them interfere. Since $A(B_H \to \psi K_S)$ and $A(B_L \to \psi K_L)$ are both $\propto \cos \beta$, the interference between them is proportional to $\cos^2 \beta$. Similarly, the interference between $A(B_H \to \psi K_L)$ and $A(B_L \to \psi K_S)$ is proportional to $\sin^2 \beta$. Obviously, a suitable linear combination of $\cos^2\beta$ and $\sin^2\beta$ will yield $\cos 2\beta$. This is why $\overset{\scriptscriptstyle(-)}{B_d} \to \psi + K \to \psi + (\pi\ell\nu)$ probes $\cos 2\beta$.[10]

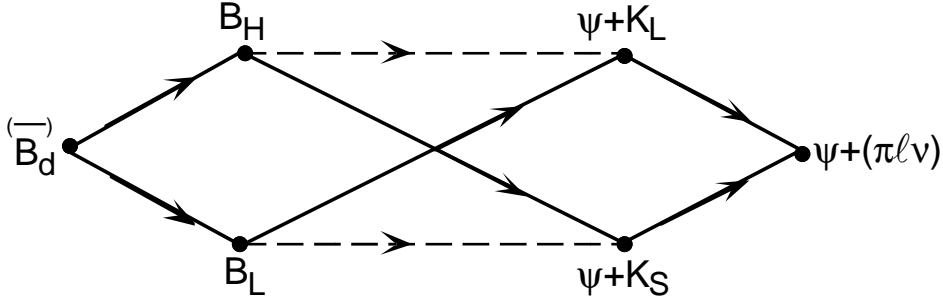

Figure 1. The paths through which the decay chain $\overset{(-)}{B_d} \to \psi + K \to \psi + (\pi \ell \nu)$ can proceed.

In conclusion, in $\overset{\scriptscriptstyle(-)}{B_d} \to \psi + K \to \psi + (\pi\ell\nu)$, the very interesting phenomenon of cascade mixing[11] makes possible the determination of $\cos 2\beta$. In combination with other measurements which probably would precede it, a measurement of the sign of $\cos 2\beta$ would eliminate all the discrete ambiguities in the CP-violating phase angles $\alpha$, $\beta$, and $\gamma$ of the Standard Model.[12]

**Acknowledgments**

This paper was written at CERN and the Max-Planck-Institut für Physik. The author warmly thanks Guido Altarelli for the excellent hospitality of CERN, and Leo Stodolsky for that of the Max-Planck-Institut and for a fruitful and enjoyable collaboration.

a discussion of the role of these signs in determining CP asymmetries, see Y. Grossman, B. Kayser, and Y. Nir, in preparation.

10. An alternative approach to probing cos 2β is described in L. Oliver, talk given at the Babar Workshop, Princeton, March 1997, and in J. Charles, A. Le Yaouanc et al., in preparation.

11. It has been suggested that cascade mixing in $\overset{(-)}{B_s} \to \Psi + K \to \psi + (\pi\ell\nu \text{ or } \pi\pi)$ might facilitate the determination of the mass difference between the mass eigenstates of the $B_s$-$\overline{B}_s$ system. See Y. Azimov and I. Dunietz, *Phys. Lett.* **B395**, 334 (1997).

12. The Moriond talk on which this written version is based included a discussion of the oscillation frequency in neutral particle mixing. This topic is treated by B. Kayser, in *ICHEP 96*, p.1135, and in B. Kayser, Stanford Linear Accelerator Center preprint SLAC-PUB-7123, so it has been omitted here.